\documentclass[final,3p,times,preprint]{elsarticle}
\usepackage[utf8]{inputenc}
\usepackage{amsfonts}
\usepackage{amsmath,amssymb}
\usepackage{natbib}
\usepackage{comment}

\usepackage{bm}
\usepackage{lineno,hyperref}
\usepackage{comment}
\usepackage{xcolor}
\usepackage{booktabs}
\usepackage{multirow}
\modulolinenumbers[5]
\usepackage[normalem]{ulem}

\begin{document}

\begin{frontmatter}


\title{Localization of peripheral reactions and sensitivity to the imaginary potential}

\author{ Imane Moumene $^{a,b}$ \fnref{IMfootnote}}  \author{Angela Bonaccorso$^{c}$  \fnref{myfootnote}}
\fntext[IMfootnote]{{imane.moumene@ced.uca.ma }}

\fntext[myfootnote]{{ bonac@df.unipi.it}}



\address{  $^a$High Energy and Astrophysics Laboratory
Department of Physics, Cadi Ayyad University, Marrakesh, Morocco.\\$^b$ Department of Physics, University of Pisa, 56127 Pisa, Italy\\
            $^c$Istituto Nazionale di Fisica Nucleare, Sezione di Pisa, Largo B. Pontecorvo 3, 56127 Pisa, Italy.}


\date{\today}

\begin{abstract}
 
 The aim of the present  study is    to make for the first time in the literature a systematic and quantitative assessment of  the evaluation of the imaginary part of the optical potential calculated within the folding model and its consequences on the localization of surface reactions. 
Comparing theoretical and  experimental reaction cross sections, for  some light projectiles on a $^9$Be target,  it has recently been shown that a single-folded s.f.  (light-) nucleus-$^9$Be imaginary optical potential  is more accurate than a  double-folded d.f. optical potential. Within  the eikonal formalism for the cross sections and phase shifts, the single-folded potential was obtained using a  n-$^9$Be phenomenological optical potential  and microscopic projectile densities. This paper is a follow-up in which we systematically study   a series of  different light and   medium-mass projectile induced reactions on $^9$Be. Our results confirm that the s.f. cross sections are larger than the d.f. cross sections and the effect increases with the projectile mass. Furthermore the strong absorption radius parameter extracted from the $S$ matrices  calculated with the  s.f. has  a stable  value $r_s$ =1.3 - 1.4 fm   for all projectile masses  in the range of incident energies  40-100AMeV.  This indicates that a clear geometrical separation  can be made between the region of surface reactions,  the region of strong absorption into other channels and the region of  weak nuclear interaction. The d.f. results are  instead much scattered and the separation between surface reactions and other channels does not seem to be consistent. Excellent  agreement  with recent experimental results confirms the  validity of our approach. \end{abstract}


\begin{keyword}
 {Exotic nuclei,  optical potentials, folding models, reaction cross sections.}
\end{keyword}
\end{frontmatter}

\section
{ Introduction}
In a seminal paper, about forty years ago, De Vries and Peng \cite{dvp} shown that the energy dependence of reaction cross sections  for medium to heavy mass nuclei could be reasonably well reproduced by using the eikonal formalism and a double folding model for the optical potential and the phase shift.  The folding model had been discussed just one year before  by Satchler and Love \cite{SL} who however warned that while the double folding model was well justified for the real potential, it was less for the imaginary potential because   the latter  must be all order in the nucleon-nucleon (nn) interaction \cite{fesh}. Indeed in their paper it was shown that the imaginary potential in some cases could be taken  to have the same shape as the real folded potential but with a renormalized strength while in other cases  a Woods-Saxon shape was taken and the parameters fitted to the elastic scattering data. The folding model was also used with good success by Kox et al. \cite{kox} in a systematic study of reaction cross sections in the intermediate energy range. From  then on the folding model has been one of the most used methods to generate optical potentials and there is a huge literature on the subject, see for example Refs.\cite{SP}-\cite{nic} and references therein.  Because of the incertitude on the method for the imaginary potential, in  calculations of elastic scattering, transfer, partial fusion and fusion  while the real potential is often obtained  by double folding the projectile and target densities with an effective nn interaction,  the imaginary part is treated phenomenologically by a Woods-Saxon potential or by  renormalizing the real folded potential. Some recent models based on complex g-matrices overcome this problem and allow also for an accurate microscopic evaluation of the imaginary part 
\cite{g-mat,g-mat2,jap}. Other authors \cite{PC} have obtained the   real potential from chiral effective field theory  and then the imaginary potential  using the dispersion relation.  Furthermore since the advent of radioactive beams, measured total reaction cross sections have been often studied by the the Glauber model and the double folding model for the imaginary potential \cite{AB0,tanih,ozawa,fui3,fui1, fui2}. Total reaction cross sections are relatively easy to measure and using the folding model one might hope to obtain information on the density distribution of the exotic nucleus projectile. On the other hand there would still be some sensitivity to the target density and the nn interaction.

$^9$Be has been one of the most used targets in reactions with radioactive beams. It is very deformed \cite{SL} and it  has strong breakup channels. There exist  an almost continuous series of neutron-$^9$Be data  as a function of the neutron incident energy.   The  optical potentials  of Ref.\cite{bobme} were able to reproduce at the same time all those data, namely the total, elastic and reaction cross sections and all available elastic scattering angular distributions. Using such potential it has recently been shown that a single-folded s.f.  (light)-nucleus-$^9$Be imaginary optical potential  is more accurate than a  double-folded d.f. optical potential. Within  the eikonal formalism for the cross sections and phase shifts, the single-folded potential was obtained using the  n-$^9$Be phenomenological optical potential \cite{bobme} and various microscopic projectile densities\cite{noi1,noi2} for light projectiles such as $^9$C,  $^8$Li and $^8$B. This paper is a follow-up in which we systematically study   a series of  different light and   medium mass projectile induced reactions on $^9$Be, thus concentrating on the projectile mass dependence rather than on the energy dependence.
 
 We hope to clarify the sensitivity of the results for reaction cross sections, $S$ matrices and strong absorption radii to the method used to obtain the optical potential and the phase shifts.  
  For all these reactions resumed in Table 1,   our interest is to assess the interaction of  the projectile with the target.  In particular we wish to understand whether there is a clear and consistent way to determine geometrical parameters that can help determine the range of impact parameters for which surface reactions dominate the projectile-target interaction from regions in which the strong absorption regime applies. For example in the case of fusion and incomplete fusion on heavy targets the experimental localization of various reactions  has been of fundamental importance in disentangling  the reaction mechanisms \cite{cook,jin}. 
  
  In the following from the  calculated $S$-matrices we obtain   the values of the strong absorption radii and then the value of the strong absorption radius parameter.  If the latter turns up to be an almost constant quantity as a function of the mass,  then  a  geometrical model such as the eikonal model, at the energies where it is  applicable, would stand on firm grounds because the region of strong and week absorption  can be clearly identified as basically independent on the nuclear masses. 
 The eikonal  approach \cite{59} is used in this paper  to obtain the  phase shifts, $S$ matrices and reaction cross sections.
   Most of the reactions  discussed here are calculated at incident energies around 60A.MeV and just a few at smaller or larger energies. At small energy our calculations for the phase-shift are performed by substituting the impact parameter with the classical distance of closest approach \cite{dvp,60}.
   
   To lend further support to our approach, at the end,  similarly to what has ben done in Refs.\cite{noi1, noi2}, we compare the energy dependence of  calculated cross sections to recent experimental values for a number of exotic nuclei.

 \section{Reminder of eikonal formulae}
 As in Ref.\cite{dvp} we calculate the
 eikonal reaction cross section according to
\begin{equation}
\sigma_{R}=2\pi\int_0^{\infty}  b~d b~ (1-\mid S_{PT}({\bf b})\mid ^2) \label{1}  \end{equation} 
where
\begin{equation}\mid S_{PT}({\bf b})\mid ^2=e^{2\chi_I(b)}\label{1bis}\end{equation} is the probability that the projectile-target (PT) scattering is elastic for a given impact parameter $\bf b$.

 The imaginary part of the eikonal phase shift is given by \begin{eqnarray}\chi_I({\bf b})&=&{1\over \hbar v}\int dz~ W^{PT}({\bf b},z)\label{2}\end{eqnarray}   
 where 
$v$ is the projectile-target velocity of relative motion and  $W^{PT}$  is  the negative defined  imaginary part of the projectile-target optical potential. We shall use two methods to obtain $W^{PT}$. One is the single folded  potential 
\begin{equation}W^{PT}_{s.f.}({\bf r})=\int  d{\bf b_1} W^{nT}({\bf b_1}-{\bf b},z) \int dz_1~\rho_P({\bf b_1},z_1). \label{4}\end{equation}
given in terms     of a nucleon-target (nT) optical potential $W^{nT}({\bf r})$, and the matter density $\rho_P({\bf b_1},z_1)$ of the projectile. In the single-folding method used in this paper, the n-target potential will be taken as  $W^{nT}_{ph}({\bf r})$, the imaginary part of the   n+$^9$Be  phenomenological nucleon-target potential  (AB) of Ref. \cite{bobme}. 

The second method defines a projectile-target potential by  the double-folding method. In this case W$^{PT}_{d.f.}$ is obtained  from  Hartree-Fock  microscopic densities $\rho_{P,T} ({\bf r})$ for the projectile and target respectively and an energy-dependent nucleon-nucleon (nn) cross section $\sigma_{nn}$, i.e., 
\begin{equation}
W^{PT}_{d.f.}({\bf r})=-{1\over 2} \hbar v \sigma_{nn}
 \int d{\bf b_1}~\rho_T({\bf b_1}-{\bf b},z)\int dz_1~\rho_P({\bf b_1},z_1). \label{5bis}\end{equation} 
Note that  Eq.\ref{5bis} can be given the same structure as Eq.(\ref{4}) by defining
\begin{eqnarray}
W^{nT}_{\rho}({\bf r})=-{1\over 2} \hbar v \sigma_{nn}\rho_T ({\bf r})\label{6}\end{eqnarray}
a single-folded zero-range  $n$-target imaginary potential. The W$^{nT}_{\rho}$ potential  of Eq.(\ref{6}) has the same range and profile as the target
density because
$\sigma_{nn}$ is a simple scaling factor. On the other hand the phenomenological potential  $W^{nT}_{ph}({\bf r})$ to be used in  Eq.(\ref{4}) has a range and in particular a profile which represents the localization of the various n-target reactions (c.f. figures in Ref.\cite{noi1,noi2}).
With the potential of Eq.(\ref{5bis}), the phase shift becomes:

\begin{equation}\chi_I({\bf b})=-{1\over 2}  \sigma_{nn}\int d{\bf b_1}~\int dz~\rho_T({\bf b_1}-{\bf b},z)\int dz_1~\rho_P({\bf b_1},z_1). \label{7}\end{equation} 

At low energy (E$_{inc}<$40A.MeV), as suggested by \cite{dvp,60,bass} we substitute the impact parameter {\bf b} with the classical distance of closest approach 
$d=b+\sqrt{b^2+a_c^2}$ where a$_c$ is the Coulomb length parameter.

A finite-range potential can also be defined as:

\begin{equation}
W^{PT}_{f.r}({\bf r})=-{1\over 2} \hbar v  \int d{\bf r_1} d{\bf r_2}~\rho_P({\bf r_1})~\rho_T({\bf r_2}) v_{nn}({\bf r_1+r-r_2} )\label{5tris}\end{equation} 
 where $v_{nn}$  can be a zero-range or a finite-range nucleon-nucleon interaction such as  Gogny   \cite{gogny} or M3Y  \cite{m3y} or a phenomenological form. However, the imaginary parts obtained in this way  need to be renormalized most of the time. For this reason we do not use such a method here.

The previous equations can be generalized in a obvious way in order to distinguish between the proton and neutron densities and the proton-neutron and proton-proton cross sections, using:
$\rho_P={\rho^n}_P+{\rho^p}_P$, and $ W^{nT}_{\rho}({\bf r})=-{1\over 2} \hbar v (\sigma_{np}{\rho^p}_T ({\bf r})+ \sigma_{pp}{\rho^n}_T ({\bf r}))$. This is the formalism followed in the present work.

The strong-absorption radius R$_s$ \cite{bass,me1} is obtained from the S-matrices calculated according to Eq.(\ref{1bis}) as the radius where $\mid S_{PT}(R_s)\mid ^2={1\over 2}$, and a "strong absorption radius parameter"  r$_s$ can be extracted from 
\begin{equation} R_s=r_s(E_{inc})(A_P^{1/3}+A_T^{1/3}).\label{rs}\end{equation}

We will in the following refer to the previous formulae, this is why, although they are well known, we have resumed them here.
In this way we have a set of quantities which define the geometry of the reactions in a transparent way and allow comparisons between different projectile-target combinations and incident energies.

In Ref.\cite{noi1, noi2} we  compared results obtained using different microscopic densities. We found that the Hartree-Fock densities reproduced better the experimental reaction cross section values. For this reason we will present here only results obtained using HF densities calculated with the code HFBTHO \cite{HF} and the Skyrme interaction  SkM* \cite{SkM*}. We have also checked that using the Skyrme interaction SLY4 \cite{Sly4} does not produce substantial differences in our findings. We use for $\sigma_{pp,np}$   the parametrization of Ref. \cite{carlos}. In this  paper we  compare results obtained with the potential defined  in Eq.(\ref{4}) and Eq.(\ref{5bis}).

  \begin{figure*}[h!]
\begin{center}
\includegraphics[scale=.5]{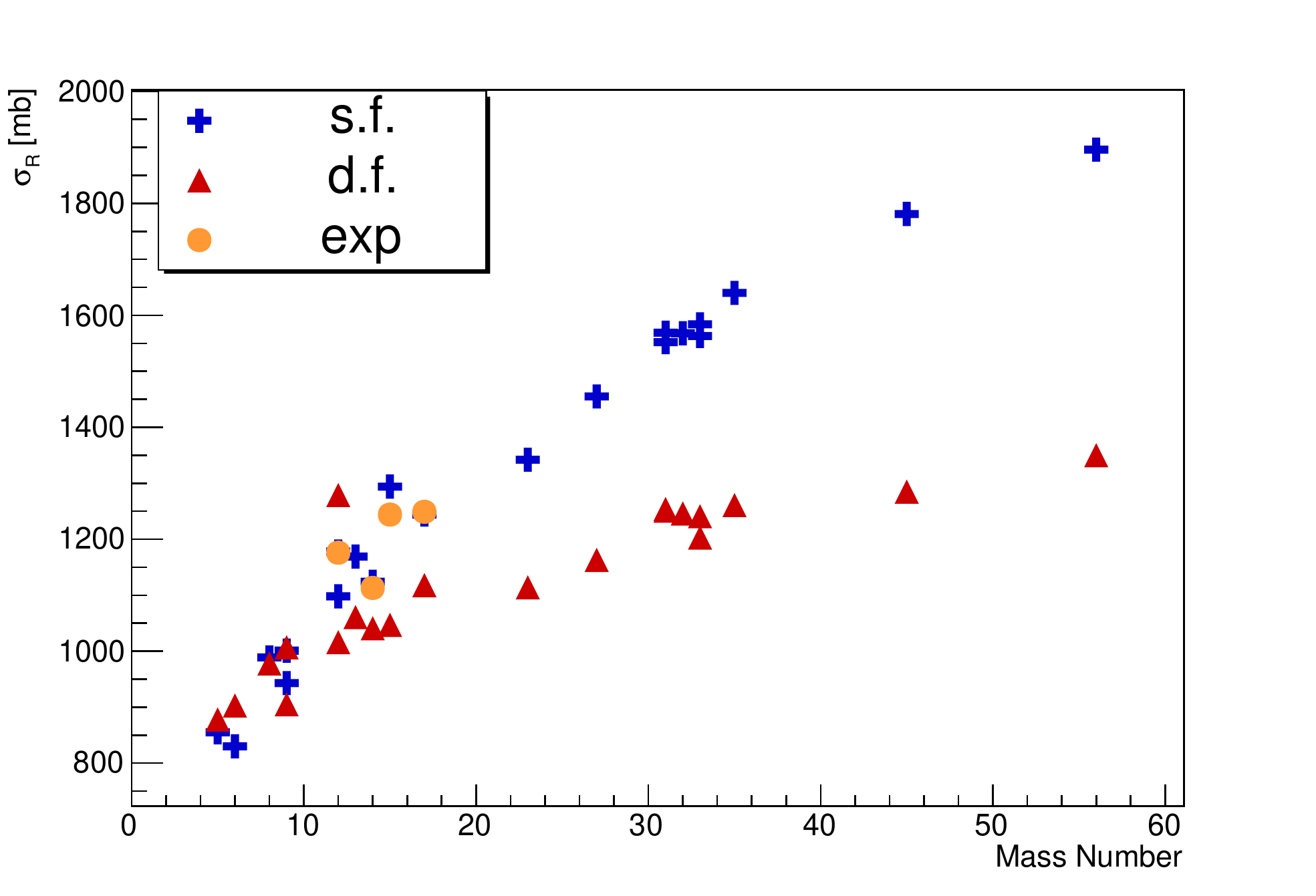}
\caption {(Color online) Reaction cross sections from Eq. (1). These include the double-folded d.f., red triangles and the single-folded s.f., blue crosses results. Orange circles for the experimental results around 60A.MeV from Table 1 and Fig.3. Mass number refers to the  A$_P$ nuclei of Table 1. }
\label{xs}
\end{center}
\end{figure*}          

\begin{figure*}
\begin{center}
\includegraphics[scale=.5]{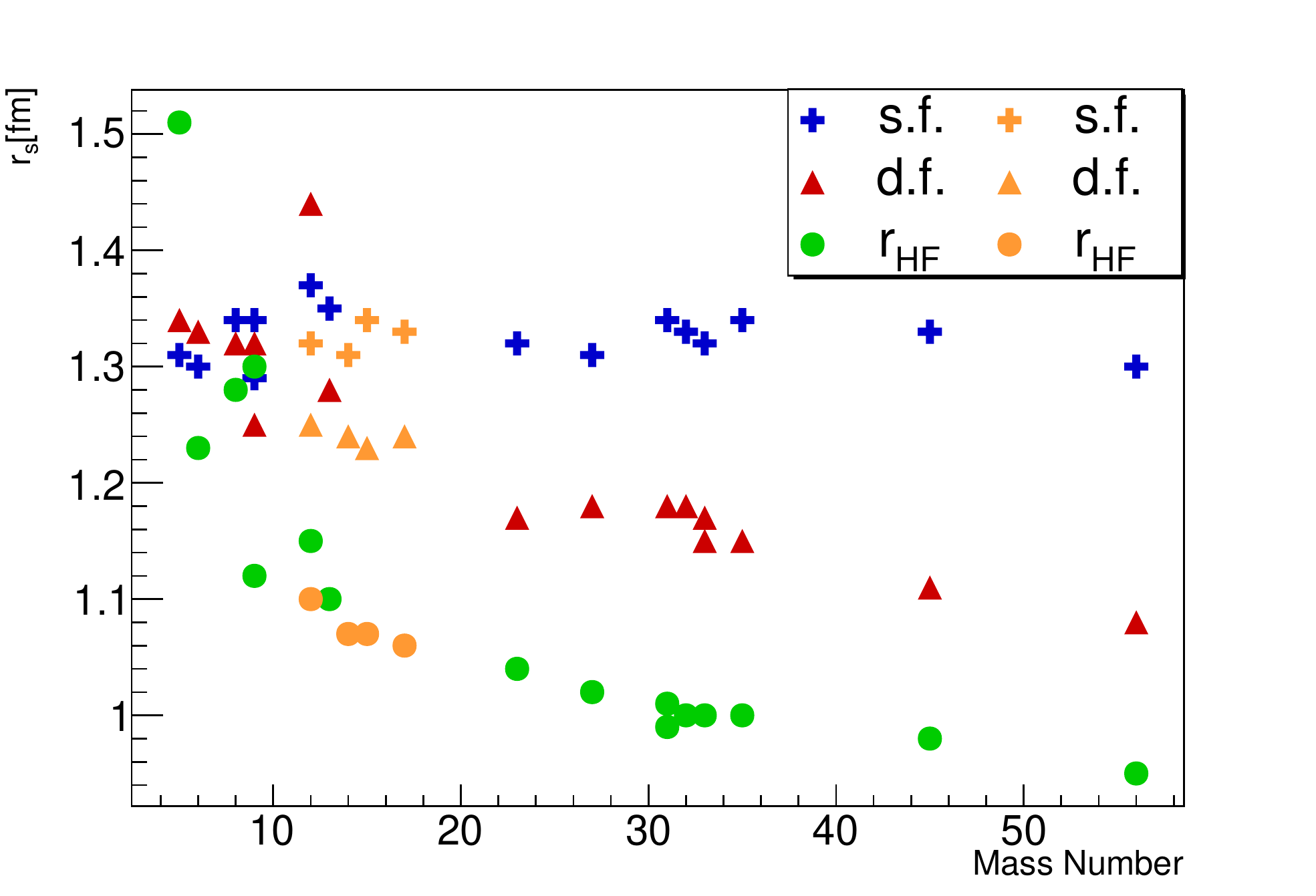}
\caption {(Color online)  Values of the r$_s$ parameters as a function of the projectile-mass. They correspond to the strong absorption radii of Eq.(\ref{rs}) from the $S$ matrices.
Red triangles from the the double-folded potentials,  and blue crosses from the single-folded potentials. Green circles are the r$_{HF}$ parameter from Table 1. Orange symbols  from the calculations relative to the data shown in Fig.3 around 60A.MeV.}
\label{rrs}
\end{center}
\end{figure*}

\begin{figure*}
\begin{center}
\includegraphics[scale=.5]{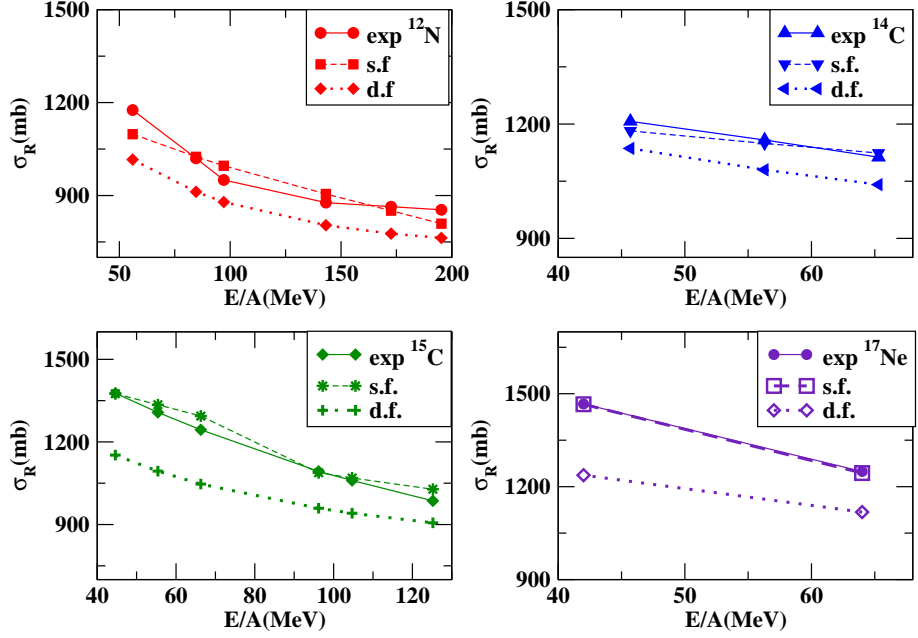}
\caption  {(Color online) Comparison of experimental cross sections and theoretical values. Red symbols $^{12}$N \cite{12N}, blue symbols $^{14}$C  \cite{14C}, green symbols $^{15}$C \cite{15C} and indigo symbols $^{17}$Ne \cite{17Ne} as indicated in the captions.  See text for details.}
\label{data}
\end{center}
\end{figure*}

\begin{figure*}
\begin{center}
\includegraphics[scale=.45]{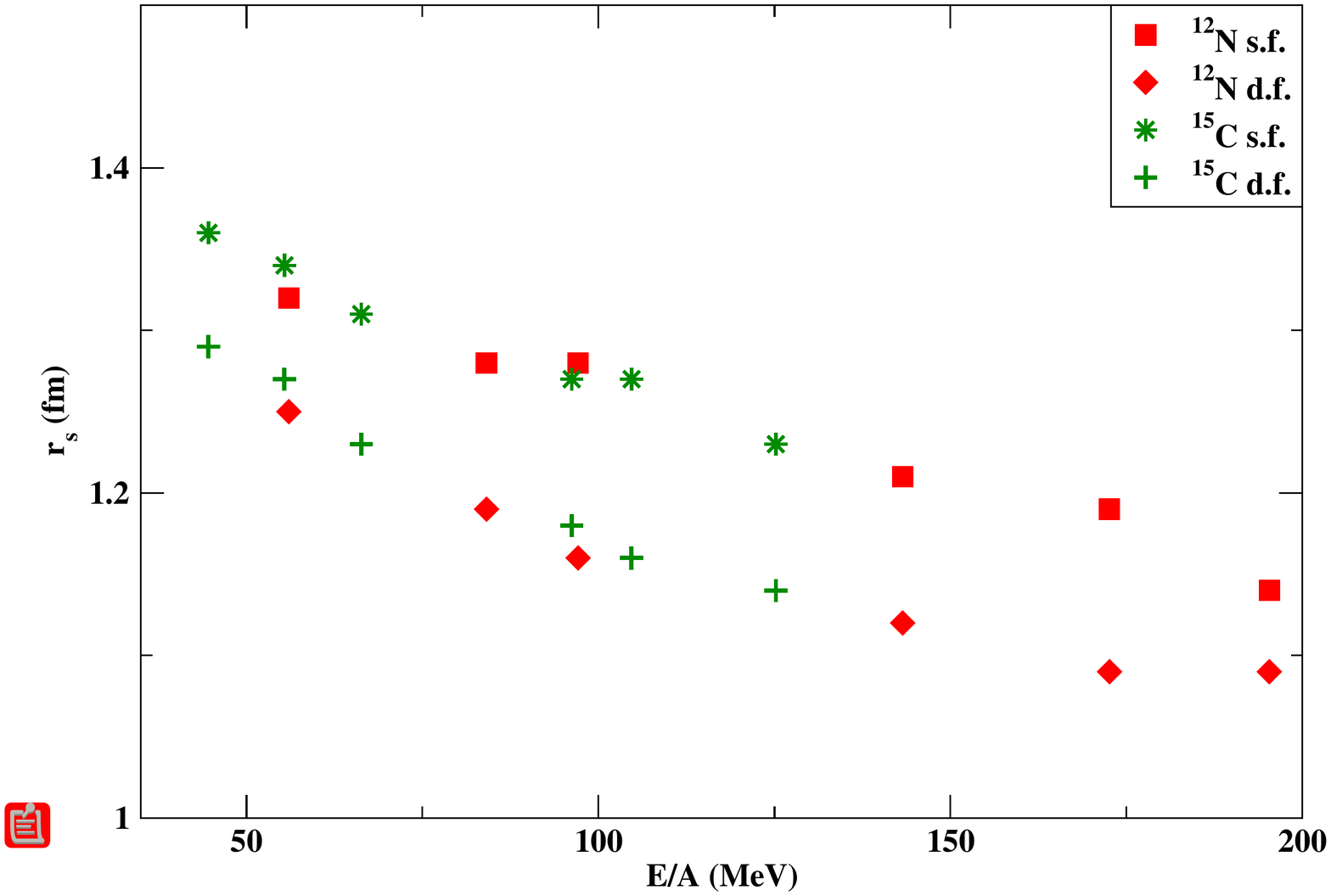}
\caption  {(Color online) Energy dependence of the strong absorption radius parameter for some the  calculations of Fig.\ref{data}: $^{12}$N, red symbols and  $^{15}$C, green symbols. See text for details.}
\label{data2}
\end{center}
\end{figure*}

\section
{ Results}

 \begin{table*} [hb]
\begin{center}
		\caption{  Reaction cross sections on $^9$Be (fourth column) of the nuclei shown in the first column, incident energies per nucleon on the second column  and strong absorption radius parameter  (fifth column) from Eq.(\ref{rs}). As for column fourth results are given for the single folding  and double folding methods. For the sake of completeness the sixth and seven columns contain the HF radii and HF radius parameter. }	
	\begin{tabular}{ccccccc}		
	\hline
	Projectile &E/A(MeV)&&$\sigma$ (mb)& $r_{s}$(fm)& $R_{HF}$ (fm)& $r_{HF}$ (fm)\\	
	\hline 
		{$^{5}${Li}} &60&{\scriptsize s.f.}&855&1.31\\
				&&	{\scriptsize d.f.}&878&1.34&2.59&1.51\\
					\hline
{$^{6}${Be}} &65.2&{\scriptsize s.f.}&830&1.30\\
	&&	{\scriptsize d.f.}&903&1.33&2.24&1.23\\
	\hline
	{$^{8}${C}}&63.8 &{\scriptsize s.f.}&989&1.34\\
	&&	{\scriptsize d.f.}&978&1.32&2.57&1.28\\
	\hline
	{$^{9}${Be}} &80&{\scriptsize s.f.}&943&1.29\\
&	&	{\scriptsize d.f.}&905&1.25&2.35&1.12\\
	\hline
			{$^{9}${C}}   &60&{\scriptsize s.f.}&1001&1.34\\
					&&	{\scriptsize d.f.}&1007&1.32&2.72&1.30\\
				
	\hline
	{$^{12}${O}}& 28.5&{\scriptsize s.f.}&1178&1.37\\
	&&	{\scriptsize d.f.}&1279&1.44&2.65&1.15\\
	\hline
		{$^{12}${N}}& 56&{\scriptsize s.f.}&1098&1.32\\
	&	&	{\scriptsize d.f.}&1016&1.25 &2.53&1.10\\
	\hline
	{$^{13}${O}} &53&{\scriptsize s.f.}&1169&1.35\\
	&&	{\scriptsize d.f.}&1061&1.28&2.60&1.10\\
	\hline
	{$^{14}${C}} & 65.3&{\scriptsize s.f.}&1124&1.31\\
	&&	{\scriptsize d.f.}&1041&1.24&2.59&1.07\\
	\hline
   {$^{15}${C}}& 66.3&{\scriptsize s.f.}&1294&1.34\\
&	&	{\scriptsize d.f.}&1047&1.23 &2.64&1.07\\
	\hline
	 {$^{17}${Ne}}& 64&{\scriptsize s.f.}&1244&1.33\\
	&	&	{\scriptsize d.f.}&1118&1.24 &2.74&1.06\\
	\hline
		{$^{23}${Si}}&85.3 &{\scriptsize s.f.}&1342&1.32\\
	&&	{\scriptsize d.f.}&1114&1.17&2.97&1.04\\
	\hline
	{$^{27}${S}} &80.7&{\scriptsize s.f.}&1455&1.31\\
	&&	{\scriptsize d.f.}&1163&1.18&3.07&1.02\\
	\hline
	{$^{31}${Ar}} &65.1&{\scriptsize s.f.}&1569&1.34\\
&	&	{\scriptsize d.f.}&1254&1.18&3.19&1.01\\
	\hline
	{$^{31}${S}}&  62.8&{\scriptsize s.f.}&1552&1.34\\
	&&	{\scriptsize d.f.}&1251&1.18&3.12&0.99\\
	\hline
	{$^{32}${Cl}}&  66.4&{\scriptsize s.f.}&1568&1.33\\
	&&	{\scriptsize d.f.}&1246&1.18&3.18&1.00\\
	\hline
	{$^{33}${Ar}}&70 &{\scriptsize s.f.}&1584&1.32\\
	&&	{\scriptsize d.f.}&1241&1.17&3.22&1.00\\
	\hline
		{$^{33}${Si}}& 73.4 &{\scriptsize s.f.}&1563&1.32\\
&	&	{\scriptsize d.f.}&1203&1.15&3.21&1.00\\
	\hline
	{$^{35}${Ca}}& 70&{\scriptsize s.f.}&1640&1.34\\
	&&	{\scriptsize d.f.}&1261&1.15&3.30&1.00\\
	\hline
	{$^{45}${Ar}}& 70&{\scriptsize s.f.}&1781&1.33\\
	&&	{\scriptsize d.f.}&1285&1.11 &3.49&0.98\\
	\hline
	{$^{56}${Ni}} & 73&{\scriptsize s.f.}&1896&1.30\\
	&&	{\scriptsize d.f.}&1350&1.08&3.65&0.95\\
	\hline
				\end{tabular}
				\end{center}
\end{table*}

Fig.\ref{xs} and Table 1 show the calculated reaction cross sections as a function of the projectile  mass. They correspond to incident energies close to 60A.MeV.
In Fig. \ref{xs} the results of Eqs.(\ref{1},\ref{4}), blue crosses, obtained by
single-folding of the (AB)-potential \cite{bobme} with the HF projectile density are almost always larger than the double - folding cross sections from Eqs.(\ref{1},\ref{5bis}), red triangles, again in agreement with what found in Ref.\cite{noi2}. Notice that the systematics presented here is different than in Ref.\cite{noi2}. There we studied the energy dependence of  $^9$C +$^9$Be scattering while here we are studying various systems  A$_p$ +$^9$Be  in the range of incident energy per nucleon  40-100AMeV. In   Table 1 we give the reaction cross sections on $^9$Be (fourth column) of the nuclei indicated in the first column, incident energies per nucleon on the second column  and strong absorption radius parameter  (fifth column) from Eq.(\ref{rs}). Results are given for the single folding  and double folding methods as indicated in the third column.The sixth and seven columns contain the HF radii of the projectiles and HF radius parameter for the sake of completeness. All calculations have been made using the Skyrme interaction SkM* \cite{SkM*}. Using the Skyrme interaction Sly4 \cite{Sly4} leads to cross sections larger by 1\% in the single folding calculations while the results are unchanged in the double folding case. In both cases the strong absorption range parameter is unchanged.

From the obtained $S$ matrices  we extract the strong absorption radii  according to $\mid S_{PT}(R_s)\mid ^2={1\over 2}$. Then using Eq.(\ref{rs})  we obtain  the values of the r$_s$ parameter given in the fifth  column of Table 1. They are also  shown  in Fig. \ref{rrs}. Red triangles are obtained from the calculations with the double-folded potentials while  blue crosses are from the single-folded potentials as a function of the projectile-mass. 
 It is very interesting to note that almost all s.f. results are concentrated in the range r$_s$ =1.3 -1.4 fm as  predicted \cite{bass} for heavy-ion reactions and always used in the past by us in analytical forms of the $S$ matrix \cite{AB0,me1} of the type 
\begin{equation}
\mid S_{PT}\mid ^2=\exp{(-ln2 e^{(R_s-b)/a})}.\label{sma}\end{equation}
The results from d.f. potentials are more scattered. This is due in part to the sensitivity to the incident energy which is treated in a more approximate way, partially to a less accurate localisation of the n-target scattering. The two red triangles corresponding to r$_s>$1.4fm  from the d.f. calculation, are due to $^5$Li and $^9$C scattering around 30A.MeV, which is the smallest energy considered in this paper. The  effect is less dramatic when using the (AB) potential and the single folding method for the target.   
Thus we note that in almost all cases the use of the phenomenological n-target potential produces  larger cross sections, cf. Fig.\ref{xs}, and a localization of the projectile-target scattering at larger impact parameters than the double folding, as seen in Fig.\ref{rrs}. This is in agreement with what found in Ref.\cite{noi2}  where we studied  the $^9$C +$^9$Be scattering. It is due to the fact that the (AB) potential contains correctly all surface effects and energy dependence of the n+$^9$Be scattering.  Note that the $^9$Be target is itself a weakly bound, strongly deformed nucleus. 
If the reaction cross sections are larger with the s.f. model it means that $\mid S_{PT}(b)\mid ^2$ is localized at larger  impact parameters and thus will be the  elastic scattering.

In Fig.\ref{rrs} and Table 1 we provide also  for comparison and completeness  the values of the radius-parameters of the HF densities,  green points, for the given projectile nuclei, defined as $R_{HF}=r_{HF} A^{1/3}$. They are scattered 
around  with respect to the projectile mass  and then we cannot extract from them a uniform  value  of the r$_{HF}$ radius parameter. This is due to the fact that the HF radius is obtained from the total density, sum of the proton and neutron densities which are very different for the nuclei studied here. 

	To give further strength to our results we present in Fig.\ref{data} a comparison of calculated and experimental  total reaction cross section data from the literature for $^{12}$Ni \cite{12N},  $^{14,15}$C  \cite{14C,15C} and $^{17}$Ne \cite{17Ne} projectiles. These results confirm that the s.f. model gives always larger cross sections and much closer to the data than the d.f. model and an excellent agreement with the energy dependence. For the two lower  energy cases in  $^{17}$Ne and $^{15}$C  we have added to our s.f. potential a small surface term of strength 0.38 and 0.40MeV  and diffusivity a=2.01fm and  a= 1.86 fm  for $^{15}$C and  $^{17}$Ne respectively.  These values are deduced from the nucleon separation energy as discussed in Refs. \cite {noi2,flome}.  This small correction to the s.f. potential is useful to take into account the projectile  nucleon breakup channel.  For completeness the relevant quantities for these nuclei, corresponding to  the cross sections at around 60A.MeV,
are shown  in Table 1 and in Figs.\ref{xs},\ref{rrs} with the orange symbols. It is very interesting to note than in both figures the results from the comparison with experimental data   fall in within the purely theoretical systematic. Finally to make a connection with   heavy-ion reaction reaction dynamics where the strong absorption regime has been widely studied and confirmed \cite{bass}, we present    in Fig.\ref{data2} the energy dependence of the strong absorption radius parameter r$_s$ as extracted from the calculations shown in Fig.\ref{data} for the two nuclei
$^{12}$Ni and ${15}$C whose data span a large energy range. As expected the strong absorption radius parameter decreases when the energy increases. However in the range 50-100AMeV its value as extracted from the single folding model is rather stable between 1.3-1.4fm. The values of r$_s$ extracted from the double folding model are all consistently smaller, between 1.2-1.3fm  in the medium energy region.
 
\section  { Conclusions} In this paper we have made for the first time in the literature a systematic comparison of calculated reaction cross sections  on a $^9$Be target. Calculations are made via a s.f.  vs. d.f. optical potential considering the strong absorption radius as a significative parameter to extract.  From the results  presented, in particular Fig. \ref{rrs} and Table  1 of this work  it appears evident  that the   d.f. method, used to calculate optical potentials, phase shifts and $S$ matrices, localises the overlap  and elastic scattering between exotic projectiles and the $^9$Be target  at smaller distances than the s.f. method and with no consistent distinction between the surface region and the region of strong absorption. This produces smaller reaction cross sections on  a $^9$Be target as can be seen in Figs.\ref{xs},\ref{data}. 
 Also it appears that  the radius parameter of the (HF) densities shows strong variations as a function of projectile mass and the difference in the number of neutrons and protons. In our opinion, this could make  it a doubtful  quantity for systematic  reaction studies.  Furthermore, because of the not consistent localization of surface reactions, results might suggest unrealistic and unphysical correlations in the analysis of experimental data.
 
  On the other hand the s.f. model has provided very stable values of the strong absorption radius parameter $r_s$=1.3 - 1.4 fm in the range of incident energies around 60 AMeV, indicating that the s.f. model is more reliable than the d.f. model in practical applications while being also better justified from a  fundamental point of view as the n-target phenomenological interaction contains all order effects. Small variations between the two values are due to the energy dependence of the cross sections. This has been elucidated by extracting the energy dependence of the radius parameter from the analysis of experimental data.  Also we suggest that a value around  $r_s$=1.4 fm could be used in Eqs. (\ref{rs},\ref{sma}) to estimate S-matrices,  predict cross sections and plan future experiments. Finally the excellent  agreement  with recent experimental results  presented in Fig.\ref{data}   and the fact that the extracted r$_s$ values agree with the purely theoretical systematic confirms the  validity  and potentiality of our s.f. approach for future  studies with rare isotope beams. 
  
 Excellent parametrizations of the n-nucleus potential exist also for heavy targets, for example the Mahaux and Sartor potential \cite{MS} or in general global parametrizations such as Koning and Delaroche \cite{KD}. The results presented here suggest that using them with an appropriate projectile density in a single folding procedure could provide more realistic and better justified potentials than those obtained from a double-folding procedure. In the case of exotic nuclei projectiles one could isolate in a clearer and more accurate way  the dependence of total reaction cross sections on the projectile density.

\section*{Acknowledgements}
We are very grateful to Prof. M. Fukuda and his group for providing us with the numerical values of their data and for allowing us to use some of them ($^{14}$C) before publication. This work was done while one of us (I.M.) was visiting the INFN, Sezione di Pisa and Department of Physics, University of Pisa.  I.M.   acknowledges the full financial support from the  University of Pisa under scheme ERASMUS+ KA107 International  Mobility. She is also grateful to Profs. M. Gaidarov and colleagues  for allowing her to run and  use results from the  code  HFBTHO\cite{HF}.

\end{document}